\documentclass[sigconf,screen]{acmart}
\AtBeginDocument{%
  \providecommand\BibTeX{{%
    \normalfont B\kern-0.5em{\scshape i\kern-0.25em b}\kern-0.8em\TeX}}}

\setcopyright{acmlicensed}
\copyrightyear{2018}
\acmYear{2018}
\acmDOI{}
\usepackage{tabularx}
\usepackage{mdframed}
\usepackage{multirow}

\acmConference[ESEM'24]{Make sure to enter the correct
  conference title from your rights confirmation email}{October 20--25,
  2018}{Barcelona, Spain}
\acmBooktitle{Proceedings of the ESEM 2024, October 20--25, 2024, Barcelona, Spain}

\begin{document}

\title{Preliminary Insights on Industry Practices for Addressing Fairness Debt}

\author{Ronnie de Souza Santos}
\affiliation{%
 \institution{University of Calgary}
  \city{Calgary}
  \state{AB}
  \country{Canada}
}
\email{ronnie.desouzasantos@ucalgary.ca}

\author{Luiz Fernando de Lima}
\affiliation{%
  \institution{CESAR}
  \city{Recife}
  \state{PE}
  \country{Brazil}
}
\email{lffpl@cesar.org.br}

\author{Maria Teresa Baldassarre}
\affiliation{%
  \institution{Università di Bari}
  \city{Bari}
  \state{BA}
  \country{Italy}
}
\email{mariateresa.baldassarre@uniba.it}

\author{Rodrigo Spinola}
\email{spinolaro@vcu.edu}
\affiliation{%
  \institution{Virginia Commonwealth University}
  \city{Richmond}
  \state{Virginia}
  \country{USA}
}

\begin{abstract}
\textbf{Context:} This study explores how software professionals identify and address biases in AI systems within the software industry, focusing on practical knowledge and real-world applications. \textbf{Goal:} We focused on understanding the strategies employed by practitioners to manage bias and their implications for fairness debt. \textbf{Method:} We employed a qualitative research method, gathering insights from industry professionals through interviews and using thematic analysis to explore the collected data. \textbf{Findings:} Professionals identify biases through discrepancies in model outputs, demographic inconsistencies, and training data issues. They address these biases using strategies such as enhanced data management, model adjustments, crisis management, improving team diversity, and ethical analysis. \textbf{Conclusion:} Our paper presents initial evidence on addressing fairness debt and lays the groundwork for developing structured guidelines to manage fairness-related issues in AI systems. \newline

\noindent \textbf{LAY ABSTRACT}. This paper explores how software professionals tackle biases in AI systems. We discovered that they identify problems by checking if the AI's outputs match real-world conditions, ensuring it performs well for different groups of people, and investigating biases in the training data. To address these issues, they use various strategies like improving the data quality, regularly updating the AI to adapt to new information, and involving a diverse range of people in the development process. Our findings provide a solid starting point for creating clear guidelines to manage these biases. These guidelines will help ensure that AI systems are not only technically accurate but also fair and equitable for everyone. This research is important for making sure that as AI technology advances, it benefits all users without reinforcing existing inequalities.

\end{abstract}


\maketitle

\section{Introduction}
The quality of software systems depends on well-informed decisions that software development teams make about the technical aspects of the system, as well as the dynamics of their work interactions \cite{caballero2023community, burge2008decision, ruhe2002software, isong2013towards}. In this context, technical debt emerges from poor technical decisions, where shortcuts or suboptimal solutions are adopted to expedite delivery, leading to future maintenance difficulties. Technical debt is defined as the accumulated cost of additional rework caused by choosing an easier, limited solution now instead of a better approach that would take longer \cite{Cunningham, li2015systematic, ernst2015measure}. Analogous to technical debt, social debt arises from poor social decisions that negatively affect the work environment. This includes ineffective communication, lack of collaboration, and unresolved conflicts, leading to stressed relationships, decreased morale, and a breakdown in teamwork. Social debt is defined as the accrued negative impact on team dynamics and productivity resulting from suboptimal social interactions and management practices ~\cite{socialDebt, tamburri2013social, kazman2019managing}.

Recently, the rapid growth of artificial intelligence has increased discussions around the importance of fairness in software engineering ~\cite{galhotra2017fairness, brun2018software, zhang2021ignorance}. This scenario has introduced a new set of decisions necessary to deliver high-quality systems. Considering the characteristics of AI-powered systems and their profound societal impact, a new type of debt has emerged within software development: fairness debt \cite{santos2024software}. While technical debt traditionally relates to the technical aspects of software implementation and social debt focuses on the human dynamics within development teams, the concept of fairness debt extends beyond these boundaries to encompass the broader societal implications resulting from suboptimal decisions and workarounds made in machine learning projects \cite{santos2024software}.

Fairness debt arises from design, implementation, or managerial choices that prioritize short-term gains at the expense of creating a context where future adjustments become costly or impractical, resulting in significant societal consequences. This differentiates software fairness debt from other types of debt. While software fairness debt can exhibit properties similar to traditional debt, such as principal and interest, its resolution typically requires a blend of technical and social strategies aimed at addressing biases in software systems. Repaying fairness debt involves activities such as auditing algorithms for bias, improving team diversity and inclusion, refining dataset management practices, adopting fairness-aware techniques, training professionals to address and recognize bias, and ensuring transparent decision-making processes \cite{santos2024software}.

Considering that discussions around fairness debt are still in their early stages and acknowledging that many strategies are being used to mitigate the effects of fairness issues caused by biased outcomes in software systems, this study presents initial findings from interviews with software professionals working on AI and machine learning projects to explore the techniques software teams are using to address fairness issues in practice. Our goal is to answer the following research question: \textbf{RQ. \textit{What strategies are industry professionals currently employing to address fairness debt in AI and machine learning projects?}} Our study makes key contributions to industry practice by identifying techniques and strategies currently used to manage fairness debt, as well as offering recommendations for improving the integration of fairness considerations into software development processes.
 

\section{Fairness Debt} \label{sec:back}
Fairness is a non-functional requirement and a critical quality attribute for software, especially for systems driven by data processes~\cite{galhotra2017fairness}. Software fairness involves the ethical principle of ensuring that software systems, algorithms, and their outcomes are equitable and unbiased across diverse groups of people, regardless of characteristics such as race, gender, ethnicity, or socioeconomic status~\cite{galhotra2017fairness}. Imbalances in fairness can arise from various sources throughout the software development lifecycle, influenced by both internal software practices and external factors. These imbalances may originate from the technologies used, the development practices employed, or the interactions among team members.

Fairness debt is defined as a collection of design, implementation, or managerial decisions that, while providing short-term benefits, establish conditions where future adjustments become costly or impractical, resulting in significant societal impacts. It involves various types of biases, including cognitive, design, historical, model, requirement, societal, testing, and training biases, all of which contribute to fairness deficiencies within software systems. The central concern of fairness debt is its impact on society, with profound implications for individuals and communities, ranging from minor inconveniences to severe societal injustices \cite{santos2024software}. 

\begin{figure}[ht]
  \centering
  \includegraphics[width=1.05\linewidth]{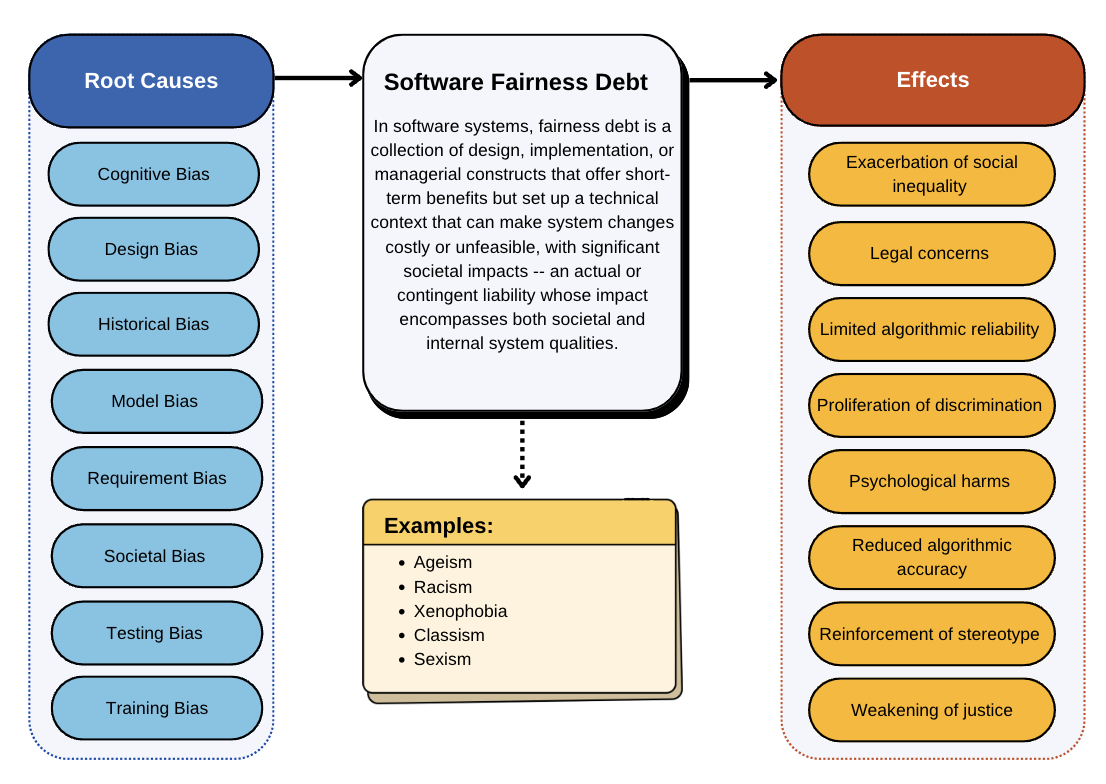}
  \caption{Conceptual of Software Fairness Debt \cite{santos2024software}}
  \label{fig:fairness_debt}
\end{figure}

Examples of fairness debt encompass a range of negative impacts, including the exacerbation of social inequalities, legal challenges, and various forms of discrimination such as ageism, classism, racism, sexism, and xenophobia. Addressing fairness debt involves a multifaceted approach that combines technical and societal perspectives. Technically, it requires the identification and mitigation of biases within data and algorithms to prevent skewed outcomes. From a societal perspective, it involves recognizing the broader implications of these biases and ensuring that software development practices promote equity and inclusivity \cite{santos2024software}. Figure 1 illustrates this concept in terms of root causes, examples, and effects in society. 

\section{Method} \label{sec:method}
In this study, we interviewed \cite{easterbrook2008selecting, seaman1999qualitative} a group of professionals involved in developing various AI and machine learning-powered software solutions. Our participants held diverse roles, including designers, software engineers, data scientists, and testers, across four distinct projects. Project A featured an application using deep learning neural networks to translate sign language into Portuguese text in real-time, enabling hearing-impaired individuals to communicate via gestures translated into text. Project B focused on digital twins and prediction models in the petroleum and energy sector. Project C involved using Large Language Models (LLMs) in educational contexts, while Project D utilized computer vision for facial recognition to identify patterns within images of individuals.

To engage professionals with diverse expertise, we employed a combination of sampling strategies \cite{baltes2022sampling}. Initially, we sent invitation to professionals working on these projects, asking those interested to participate in the study and provide their preferred dates and times for interviews, using convenience sampling. We then employed snowball sampling by asking participants to suggest other professionals who might be interested in joining the study. To further refine our sample, we incorporated theoretical sampling by sending direct invitations to individuals we suspected could provide valuable insights into the understanding of the problem. This approach enabled us to capture a wide range of perspectives and experiences from professionals.

\subsection{Data Collection}
During data collection, interviews were conducted in three rounds, starting with seven participants per round and increasing the number in the final round as responses started to reach saturation, e.g., become repetitive. Hence, between June 1 and July 5, 2024, we interviewed a total of 22 professionals. Interviews were conducted online following a pre-established guide, with rounds 1 and 2 including 7 participants each and round 3 comprising 8 participants. The interviews ranged from 23 to 42 minutes, over 3 hours of recorded audio. Three participants could not participate in recorded interviews, so they completed a questionnaire with open-ended questions.

\subsection{Data Analysis}
For data analysis, we employed thematic analysis \cite{cruzes2010synthesizing}, a method used to identify and analyze patterns within qualitative data. This approach is widely used in software engineering research and helps identify cross-references among different data sources. After each round of interviews, we systematically reviewed the transcripts to identify key elements that could answer our research question. This iterative process involved coding the data, categorizing the codes into themes, and refining these themes as new data emerged. We continued this process until we achieved a good amount of evidence. This structured approach allowed us to gather detailed information, synthesize narratives from different participants, and draw conclusions to provide clear and actionable insights for practitioners.

\section{Findings} \label{sec:findings}
We interviewed 22 software professionals using convenience, snowball, and theoretical sampling methods. These professionals are actively involved in developing AI-powered systems such as deep learning neural networks, prediction models, LLMs, and computer vision for facial recognition. The participants hold diverse roles in software development, including data scientists, programmers, software QA/testers, designers, and software project managers. Additionally, recognizing the critical role of diversity in addressing bias and fairness debt, we included individuals from equity-deserving groups, encompassing non-male professionals, individuals with disabilities, non-white individuals, LGBTQIA+ individuals, and neurodivergent individuals. Understanding their experiences and perspectives is essential for discussing bias in AI development. Table \ref{tab:Demographics} summarizes the information of our group of participants, and our two main findings are presented below.

\begin{table}
\centering
\caption{Demographics}
\renewcommand{\arraystretch}{1}
\label{tab:Demographics}
\begin{tabular}{llr}
\toprule
\multirow{2}{*}{ \textbf{Gender} } 
& Men & 15\\
& Women & 6 \\ 
& Non-binary & 1 \\ \midrule 

\multirow{5}{*}{ \textbf{Role} } 
& Data Scientists & 9 \\
& Designers & 4 \\
& Programmers & 3 \\
& Testers & 3 \\
& Researcher & 2 \\
& Managers & 1 \\ \midrule 

\multirow{4}{*}{ \textbf{Education} } 
& Bachelor's Degree & 8 \\ 
& Postbaccalaureate & 3 \\ 
& Master's Degree & 6 \\ 
& PhD Degree & 5 \\ \midrule 

\multirow{4}{*}{ \textbf{Experience} }  
&1-3 years & 3 \\ 
&3-5 years & 8 \\ 
&5-10 years & 6 \\ 
&More than 10 years & 5 \\ \midrule 

\multirow{3}{*}{ \textbf{Ethnicity} } 
& White & 19 \\
& Mixed-race & 2 \\
& Black & 1 \\ \midrule 

\multirow{2}{*}{ \textbf{Disability} } 
& Without & 20\\
& With & 2 \\ \midrule 

\multirow{3}{*}{ \textbf{LGBTQIA+} } 
& No & 17 \\
& Yes & 5 \\ \midrule 

\multirow{3}{*}{ \textbf{Neurodivergent} } 
& No & 18 \\
& Yes & 4 \\
\bottomrule
\end{tabular}
\flushleft
\end{table}

\subsection{How Do Software Professionals Identify Bias in AI Projects?}
Our analysis reveals that software professionals use a range of strategies to identify biases in AI systems. These strategies primarily involve searching for discrepancies between model outputs and real-world conditions, including evaluating the correctness and variability of outputs across demographic groups and assessing the presence of biased training data. Additionally, professionals recognize bias during data collection and preparation, as well as before releasing the system by evaluating algorithmic behavior through testing datasets. More specifically, we translated these strategies into the following practices:

\begin{itemize}
    \item \textbf{Mismatch Between Model Output and Reality}. Biases are often detected when professionals notice that their algorithms produce results that significantly deviate from what is expected in real-world situations. This occurs when the model's output does not align with actual conditions or practical realities, indicating that the model may have difficulty generalizing from its training data to real-world applications. Such mismatches can reveal underlying issues with how the model processes and interprets data, suggesting that the model might not fully capture the complexities or nuances of the real-world scenarios it is intended to address.
    
    \item \textbf{Inconsistent Performance Across Different Demographics}. Biases become evident when a model demonstrates varying levels of performance or accuracy across different demographic groups, such as age, gender, ethnicity, or socio-economic status. For instance, a model might perform exceptionally well for one group but poorly for another. This inconsistency indicates that the model may be inadvertently favoring certain groups over others, which could be a sign of biased training data or unequal representation in the dataset. Identifying such discrepancies helps in understanding and addressing potential biases embedded in the model's design and training process.
    
    \item \textbf{Dependence on Biased Training Data}. Biases are identified when there is clear evidence that the training data used to develop the model contains inherent biases or is not representative of the entire population. For example, if the training data overrepresents certain groups while underrepresenting others, the model is likely to reflect and perpetuate these biases in its predictions and decisions. This dependence on biased training data can lead to skewed outcomes and reinforce existing inequalities, making it important to ensure that training data is diverse, representative, and free from bias to develop fair and effective models.
    
    \item \textbf{Algorithmic Behavior During Testing}. Biases can be detected by analyzing how the model behaves during the testing phase, where it is evaluated against different test scenarios. Professionals may observe that the model’s performance varies inconsistently across these scenarios, revealing potential biases in how the model handles different types of data or situations. Such inconsistencies during testing can highlight underlying biases in the model's design or training, such as poor handling of underrepresented cases or specific conditions. Understanding these behavioral patterns reveals the importance of a robust testing process, which is essential for identifying and addressing biases to ensure that the model performs fairly and effectively across a wide range of scenarios.
\end{itemize}





These findings show that software professionals adopt diverse approaches to uncover biases in AI systems. They emphasized that the perspective of bias can emerge from various sources, not just the exploration of data. This includes discrepancies between model outputs and real-world conditions, inconsistent performance across different demographic groups, and biases observed during algorithmic testing. Examples of evidence collected from interviews to illustrate these findings are presented in Table \ref{tab:bias_identification}

\begin{table}[h!]
\caption{Bias Identification}
\centering
\label{tab:bias_identification}
\begin{tabularx}{\linewidth}{p{2.4cm} X}
\toprule
\textbf{Identification} & \textbf{Evidence Examples} \\ \hline 

\textbf{Mismatch \newline Between Model \newline Output and \newline Reality} & \textit{``The main issue that ends up being problematic is the type of response it brings when the model provides an answer that really doesn't match reality.''} (P1) \newline \\ 
& \textit{``When the model brings a result more focused on a specific domain, it always responds according to the training data (instead of reality).''} (P7) \\
\hline

\textbf{Inconsistent \newline Performance \newline Across Different \newline Demographics} & \textit{``We identified that for some people, the algorithm performed better based on the regions of interest, such as hand movements.''} (P6) \newline \\ 
& \textit{``We realized the importance of diversity when our model only recognized actions performed by white people.''} (P10) \\
\hline

\textbf{Dependence on \newline Biased Training \newline Data} & \textit{``We tried to avoid bias by balancing the data... but in reality, we don't always have balanced data.''} (P5) \newline \\ 
& \textit{``We noticed that certain outcomes were repeatedly skewed, leading us to investigate the training dataset and uncover inherent biases.''} (P8) \\
\hline
\textbf{Algorithmic \newline Behavior During 
\newline Testing} & \textit{``The testing phase is extremely important to try to identify biases in algorithms. Diversifying the test dataset is key.''} (P17) \newline \\ 
& \textit{``(during testing) bias often comes from insufficient samples that do not represent the total population.''} (P14) \\
\bottomrule
\end{tabularx}
\end{table}

\subsection{How Do Software Professionals Address Bias in AI Projects?}
Our analysis reveals that software professionals employ a variety of strategies to address biases identified in AI system projects post-release. These strategies encompass different aspects of managing and mitigating bias to enhance model performance and fairness. More specifically, we have categorized these strategies as follows:

\begin{itemize}
    \item  \textbf {Data Enrichment}. This strategy focuses on enhancing and diversifying the dataset to better reflect real-world diversity and minimize biases. It involves several key activities, including data cleaning and preprocessing to remove or correct harmful biases within the existing data, acquiring additional data sources to cover different scenarios, and conducting preliminary bias analysis to identify and address potential biases before training the model. By creating a more balanced and representative dataset, this practice helps develop fairer and more effective models.
    \item \textbf {Model Adjustment}. This strategy centers on continuously assessing and refining the model to manage emerging biases and maintain performance across different conditions. It includes regular model testing with diverse scenarios to detect performance variations and potential biases, retraining the model with new, diverse data to adapt to evolving conditions, and implementing thorough model audits to ensure transparency, accountability, and adherence to fairness standards. This practice ensures the model remains effective and equitable as new data and scenarios are encountered.
    \item \textbf{Crisis Response}. This strategy involves establishing protocols for managing and communicating about issues that arise during or after the deployment of the AI system. It includes creating clear procedures for addressing and resolving biases or other problems and involving users in the process by keeping them informed about potential issues. These measures help manage unexpected problems and maintain user trust and confidence in the system.
    \item \textbf{Diverse Team Integration}. This strategy emphasizes the importance of including individuals from varied backgrounds within the software development team to provide a broad range of perspectives. A diverse team ensures that various viewpoints and experiences are considered in the development process. This approach contributes to more comprehensive problem-solving and helps in recognizing potential biases more effectively.
    \item \textbf{Ethical and Fairness Review}. This strategy involves integrating ethical considerations and systematic bias assessments throughout the software development process. It includes incorporating ethical guidelines to ensure the model adheres to fairness and integrity principles and continuously evaluating the model for biases at various stages to address and mitigate them proactively. This practice ensures the software meets ethical standards throughout its development and deployment.
\end{itemize}

The diverse strategies used by software professionals to tackle identified biases in AI systems highlight the complexity and multifaceted nature of bias management. From data refinement and model adjustments to crisis management and ethical analysis, each strategy targets different aspects of bias mitigation. By employing this range of strategies, professionals aim to ensure that AI solutions are not only technically sound but also socially responsible. Table \ref{tab:bias_mgmt} illustrates these findings with quotations extracted from our participants.

\begin{table}[h!]
\caption{Bias Management}
\centering
\label{tab:bias_mgmt}
\begin{tabularx}{\linewidth}{p{2cm} X}
\toprule
\textbf{Strategy} & \textbf{Evidence Examples} \\ \hline 
\textbf{Data \newline Enrichment} & \textit{``If we identify something bringing a biased response or risk, we map it and seek to understand why the model is doing that.''} (P1) \newline \\
& \textit{``There are techniques for detoxification... removing aggressive language from data.''} (P7) \\
\hline

\textbf{Model \newline Adjustment} & \textit{``Techniques for Concept Drift attempt to identify out-of-pattern situations and adjust the model.''} (P13) \newline \\
& \textit{``So, to solve a bias in this approach, you would need to adjust the model, but mainly seek more diverse data.''} (P10) \\
\hline

\textbf{Crisis \newline Response} & \textit{``If it happens, it becomes much more of a crisis management situation, meaning making it very clear that the solution is in development.''} (P4) \newline \\

& \textit{``Because when this problem was happening a lot,  the team would try to go back to the literature to explore other models and solutions. But the client didn’t really understand this, so they would often ask, 'Okay, but what about the software?'.''} (P4) \\
\hline

\textbf{Diverse \newline Team \newline Inclusion} & \textit{``It’s important to have people from different groups involved, as it ensures empathy and more comprehensive problem-solving.''} (P5) \newline \\
& \textit{``Building diverse teams brings a variety of experiences that contribute to a more inclusive solution.''} (P10) \\
\hline

\textbf{Ethical and \newline Fairness \newline Review} & \textit{``Incorporating ethical analysis into the development process helps ensure the model aligns with fairness standards.''} (P11) \newline \\
& \textit{``The development of white-box algorithms aims to provide transparency and ethical scrutiny.''} (P13) \\

\bottomrule
\end{tabularx}
\end{table}

\section{Discussions} 
Fairness debt represents the long-term costs associated with design, implementation, or managerial decisions that offer short-term advantages but ultimately create conditions where future corrections are difficult or costly. Effectively addressing fairness debt requires recognizing and mitigating biases that originate from various sources, including cognitive, design, historical, model, requirement, societal, testing, and training biases. Understanding these root causes is vital for managing and alleviating fairness debt \cite{santos2024software}.

Our findings indicate that while professionals in our study recognized several sources of bias, such as cognitive, testing, and training, not all the root causes identified in the literature were explicitly mentioned. Specifically, biases related to historical and requirements were less frequently highlighted. This suggests that although professionals have a broad understanding of bias sources, there remains an opportunity for deeper exploration and comprehension of all aspects of fairness debt.

Furthermore, software professionals apply a range of strategies to address biases, demonstrating a comprehensive approach to managing fairness debt. Their practices span from refining data and adjusting models to managing crises and including diverse perspectives. These strategies aim not only to improve the technical performance of AI systems but also to address the ethical and societal implications of bias.

Answering our RQ, industry professionals employ strategies that can be leveraged to contain the effects of fairness debt in AI and machine learning projects. They focus on enhancing data management through practices like refining datasets, mitigating harmful biases, and acquiring additional data to address gaps. Model management practices include regular testing with diverse scenarios, retraining to adapt to new data, and ensuring transparency through auditing. Effective crisis management involves establishing protocols for addressing and communicating issues as they arise. Additionally, professionals emphasize the importance of including diverse perspectives within development teams and integrating ethical analysis throughout the development process.

These preliminary findings are significant in laying the groundwork for developing detailed guidelines for managing fairness debt. Just as technical debt literature provides established practices for dealing with technical challenges \cite{brown2010managing, falessi2013practical, lenarduzzi2021systematic}, these insights offer the first step toward creating structured methods for addressing fairness debt. Formulating such guidelines will be essential for ensuring that software systems are not only technically proficient but also socially responsible and equitable.

\section{Conclusions} 
\label{sec:conclusions}
Our study provides valuable insights into how software professionals identify biases in AI systems. We found that professionals focus on detecting various types of biases, including those arising from discrepancies between model outputs and real-world conditions, inconsistencies across demographic groups, and inherent biases in training data. These findings highlight the importance of not only recognizing but also understanding the root causes of bias. By concentrating on specific indicators, such as the alignment between model outcomes and reality and demographic performance variability, professionals are better equipped to identify potential fairness issues early in the development process. This proactive identification helps in addressing biases before they affect the system’s performance and fairness.

In terms of bias management, we highlighted a range of strategies that professionals use to address and mitigate identified biases. These include technical strategies, such as refining data management practices and adjusting models regularly, as well as human-related strategies, such as integrating diverse perspectives within development teams and effectively managing crises. Looking at these results, we conclude that managing bias requires a balanced approach that combines both technical solutions and human insights to ensure that AI systems are not only accurate but also fair and equitable. 

Finally, relating this study to the research on fairness debt, our findings demonstrate the need for a more structured approach to managing fairness-related issues and their effects on society. By highlighting the practical strategies currently employed in the industry, we provide the initial foundation for developing guidelines to address fairness debt. We propose that future work should focus on creating these guidelines, building on the insights gathered from this study, and exploring how these strategies can be refined and standardized to enhance both the technical and ethical dimensions of software fairness.

\bibliographystyle{ACM-Reference-Format}
\bibliography{bib}

\end{document}